\documentstyle[12pt]{article}

\makeatletter
\@addtoreset{equation}{section}
\makeatother
\renewcommand{\theequation}{\thesection.\arabic{equation}}

\newcommand{\journal}[4]{{\em #1~}#2\,(19#3)\,#4;}

\newcommand{\hpa}{\journal {Helv. Phys. Acta}}

\newcommand{\ijmp}{\journal {Int. J. Mod. Phys.}}

\newcommand{\pr}{\journal {Phys. Rev.}}

\newcommand{\rmp}{\journal {Rev. Mod. Phys.}}
\newcommand{\cmp}{\journal {Comm. Math. Phys.}}

\newcommand{\np}{\journal {Nucl. Phys.}}
\newcommand{\pl}{\journal {Phys. Lett.}}

\newcommand{\prep}{\journal {Phys. Rep.}}

\newcommand{\yf}{\journal {Yad. Fiz.}}

\newcommand{\annp}{\journal {Ann. Phys. (N.Y.)}}

\makeatletter
\@addtoreset{equation}{section}
\makeatother
\renewcommand{\theequation}{\thesection.\arabic{equation}}

\renewcommand{\a}{\alpha}
\renewcommand{\b}{\beta}
\newcommand{\g}{\gamma}           
\renewcommand{\d}{\delta}         \newcommand{\D}{\Delta}
\newcommand{\e}{\varepsilon}
\newcommand{\k}{\kappa}
\renewcommand{\l}{\lambda}        
\newcommand{\m}{\mu}
\newcommand{\n}{\nu}
\renewcommand{\o}{\omega}         \renewcommand{\O}{\Omega}
              
\newcommand{\r}{\rho}
\newcommand{\s}{\sigma}           \renewcommand{\S}{\Sigma}
         
\newcommand{\f}{{\phi}}           

\newcommand{\x}{\xi}

\renewcommand{\AA}{{\cal A}}

\newcommand{\CC}{{\cal C}}

\newcommand{\FF}{{\cal F}}
\newcommand{\GG}{{\cal G}}
\newcommand{\HH}{{\cal H}}

\newcommand{\LL}{{\cal L}}
\newcommand{\MM}{{\cal M}}

\newcommand{\PP}{{\cal P}}

\newcommand{\SS}{{\cal S}}

\newcommand{\VV}{{\cal V}}

\newcommand{\Sla}{\raise.15ex\hbox{$/$}\kern -.70em D}

\newcommand{\lp}{\left(}\newcommand{\rp}{\right)}
\newcommand{\lc}{\left[}\newcommand{\rc}{\right]}
\newcommand{\lac}{\left\{}\newcommand{\rac}{\right\}}

\newcommand{\complex}{{\kern .1em {\raise .47ex
\hbox {$\scriptscriptstyle |$}}
    \kern -.4em {\rm C}}}
\newcommand{\real}{{{\rm I} \kern -.19em {\rm R}}}
\newcommand{\rational}{{\kern .1em {\raise .47ex
\hbox{$\scripscriptstyle |$}}
    \kern -.35em {\rm Q}}}
\renewcommand{\natural}{{\vrule height 1.6ex width
.05em depth 0ex \kern -.35em {\rm N}}}

\newcommand{\half}{\frac 1 2}
\newcommand{\pa}{\partial}

\newcommand{\ie}{{{\em i.e.} }}

\newcommand{\sla}{\raise.15ex\hbox{$/$}\kern -.57em}
\newcommand{\twiddle}{\lower.9ex\rlap{$\kern -.1em\scriptstyle\sim$}}

\newcommand{\equ}[1]{(\ref{#1})}
\newcommand{\eq}{\begin{equation}}
\newcommand{\eqn}[1]{\label{#1}\end{equation}}
\newcommand{\eea}{\end{eqnarray}}
\newcommand{\eqa}{\begin{eqnarray}}
\newcommand{\eqan}[1]{\label{#1}\end{eqnarray}}
\newcommand{\ba}{\begin{array}}
\newcommand{\ea}{\end{array}}
\newcommand{\eqac}{\begin{equation}\begin{array}{rcl}}
\newcommand{\eqacn}[1]{\end{array}\label{#1}\end{equation}}

\renewcommand{\=}{&=&} 
\renewcommand{\in}{\int d^4x}
\newcommand{\ds}[1]{\frac{\delta\Sigma}{\delta {#1}}}

\newcommand{\dd}[1]{\frac{\delta}{\delta {#1}}}
\newcommand{\pp}[1]{\frac{\partial}{\partial {#1}}}
\newcommand{\non}{\nonumber \\}
\newcommand{\6}{\partial}
\newcommand{\as}{\,{}^\ast\!}
\newcommand{\pam}{{\partial_\mu}}
\newcommand{\pan}{{\partial_\nu}}

\newcommand{\dxm}{dx^\mu }
\newcommand{\dxn}{dx^\nu }
\newcommand{\dxr}{dx^\rho }

\renewcommand{\title}[1]{\null\vspace{25mm}

\noindent{\Large{\bf #1}}\vspace{10mm}

\noindent {\large By }}
\newcommand{\authors}[1]{\noindent{\large #1}\vspace{3mm}

}
\newcommand{\address}[1]{\noindent #1\vspace{5mm}

}
\renewcommand{\abstract}[1]{\vspace{10mm}

\noindent{\small{\em Abstract.} #1}\vspace{2mm}

} 

\begin{document}


\hspace*{\fill} REF. TUW 96-18

\title{The finiteness of the four dimensional antisymmetric tensor 
field model in a curved background}
\authors{U. Feichtinger, O. Moritsch\footnote{Work supported in part by the
         ``Jubil\"aumsfonds der \"Osterreichischen Nationalbank''
         under Contract Grant Number 5393.}, 
         J. Rant\footnote{Work supported in part by the
         ``Fonds zur F\"orderung der Wissenschaftlichen Forschung''
         under Contract Grant Number P11354-PHY.},
         M. Schweda and H. Zerrouki\footnote{Work supported in part by the
         ``Fonds zur F\"orderung der Wissenschaftlichen Forschung''
         under Contract Grant Numbers P10268-PHY and P11582-PHY.}}
\address{Institut f\"ur Theoretische Physik, 
         Technische Universit\"at Wien\\
         Wiedner Hauptstra\ss e 8-10, A-1040 Wien (Austria)}         

\centerline{}
\centerline{}
\leftline{November 1997}
\centerline{}
\centerline{}

\abstract{
A renormalizable rigid supersymmetry for the four dimensional
antisymmetric tensor field model in a curved space--time background
is constructed. A closed algebra between the BRS and the supersymmetry
operators is only realizable if the vector parameter
of the supersymmetry is a covariantly constant vector field. This 
also guarantees that the corresponding transformations lead to a genuine
symmetry of the model. \\
The proof of the ultraviolet finiteness to all orders of perturbation theory
is performed in a pure algebraic manner by using 
the rigid supersymmetry.
}


\setcounter{page}{0}
\thispagestyle{empty}


\newpage
\section{Introduction}

One of the most exciting investigations of the last decade was the 
study of certain problems arising in gauge theory, which led
to important developments and deep insights into the topology and 
geometry of low dimensional manifolds. A well--known example is
the analysis of topological invariants~\cite{hor1,blau1} 
of four dimensional manifolds by Donaldson~\cite{don1,don2}. 
Another contribution was given by Witten~\cite{wit1}, namely
the construction of the so--called topological Yang--Mills model on a four 
dimensional manifold. After that many such topological
models, like Chern--Simons theory, BF models and others have been
discussed and many new features, as the description of
invariants of knots in terms of the Chern--Simons theory~\cite{wit2},
have been found.

The main property of topological field theories~\cite{birm1} 
is that the observables only depend on the global structure of the 
space--time manifold on which the model is defined. Particularly this
implies that these quantities are independent of any metric which
may be used to construct the classical theory. There exist two 
different types of topological field theories, namely the Witten--type
models and the Schwarz--type models. The first one is characterized
by the fact, that the whole gauge fixed action can be written as a
BRS variation, whereas for the second one only the gauge--fixing part
of the action is given by a BRS variation. The most famous example of
a Witten--type model is the topological Yang--Mills theory and
representatives of the Schwarz--type models are the Chern--Simons theory
and the BF models.

In particular, the topological BF models describe the coupling of an 
antisymmetric tensor field to the Yang--Mills field strength. 
Chronologically, such models have been first used in interacting 
string theories and nonlinear sigma 
models~\cite{ogi1,des1,hay1,kalb1,crem1,nam1,freed1}. 
Their topological nature has been analyzed much later~\cite{hor1,blau1}.
Furthermore, these models are also studied because of their connection
with lower dimensional quantum gravity. Especially, the 
Einstein--Hilbert gravity in three space--time dimensions, with
and without cosmological constant, can be naturally formulated
in terms of the BF models~\cite{birm1,wit3,sor1}.

In general, it is known that the BF models, due to the presence 
of zero modes, require a highly nontrivial 
quantization~\cite{mieg1,alvi1,dia1,sor2,abu1}, which implies
several ghost generations for the gauge--fixing procedure. 
This can be done in an elegant manner by using the Batalin--Vilkovisky
quantization procedure~\cite{bat1}.

The aim of this work is to analyze the perturbative finiteness of
the four dimensional BF model. We generalize the discussion already
carried out in the flat space--time limit~\cite{sor2} and take into
consideration the presence of a curved background. 
For this purpose, we will use the concept
of the extended BRS symmetry~\cite{pig1} and we will follow the way 
demonstrated for the Chern--Simons theory~\cite{pig2}. 
A renormalizable rigid supersymmetry~\cite{pig2,zer1,zer2} will play 
an important role which, in the flat space--time limit, is a common 
feature of many topological field theories~\cite{del1,birm2,luc1,wer1,pig3}. 
We will see that
the algebra between the BRS operator and the generators of translation
and rigid supersymmetry closes on--shell. Furthermore, the closure
of the algebra also requires a constraint for the corresponding
infinitesimal supersymmetry parameter.
This fact limited our analysis to be only valid for a curved manifold
admitting a gradient vector.

Our present work is organized as follows: in Section 2 we give
an overview concerning the classical algebraic properties of the four
dimensional BF model. Next, we construct the rigid supersymmetry
and analyze the off--shell algebra.
In Section 3 we will discuss the stability of the model by using 
cohomology techniques, and see that the symmetries do not allow 
any deformation of the classical action.
The last section is devoted to the study of anomalies, which will
complete our proof of the perturbative finiteness. 
Some details concerning the trivial counterterms can be found in the
final appendix.


\section{The classical BF model}  \label{sec1}

The BF models can be defined on manifolds $\MM$ in
arbitrary dimensions $(n+2)$, with a gauge group $G$, according 
to~\cite{hor1,blau1,birm1,pig3}
\eq
\SS_{BF} = Tr \int_{\MM} BF = \frac{1}{2n!} Tr \int d^{\,n+2}x\,
\e^{\m_1\cdots\m_{n+2}} B_{\m_1\cdots\m_{n}}F_{\m_{n+1}\m_{n+2}} \ ,
\eqn{bf-general}
where the two--form
\eq
F=dA+\half [A,A] = \half F_{\m\n} dx^\m dx^\n
\eqn{f-2form}
is the Yang--Mills field strength of the gauge connection one--form
$A=A_\m dx^\m$ and the field 
$B=\frac{1}{n!}B_{\m_1\cdots\m_{n}}dx^{\m_1}\cdots dx^{\m_n}$ 
is a $n$--form. Of course, this action being metric independent
has a topological character~\cite{hor1,blau1}.

\subsection{The four dimensional BF model in flat space--time}

In terms of differential forms we start with the topological invariant
classical action on an arbitrary space--time four--manifold $\MM$
\eq
S_{inv}= Tr \int_\MM BF = \int_\MM B^a F^a \ ,
\eqn{top-bf}
where the two--forms for the antisymmetric tensor field $B^a$ and
the field curvature $F^a$ are given by
\eqa
B^a \= \half B^a_{\mu\nu} dx^\mu dx^\nu \ , \non  
F^a \= \half F^a_{\mu\nu} dx^\mu dx^\nu 
    = \half (\6_\mu A^a_\nu - \6_\nu A^a_\mu 
    + f^{abc} A^b_\mu A^c_\nu) dx^\mu dx^\nu \ ,
\eqan{two-forms}
with the gauge field $A^a_\mu$. The fields belong to the adjoint 
representation of the gauge group $G$, assumed to be 
compact and semi--simple\footnote{Gauge group indices
are denoted by Latin letters $(a,b,c,...)$ and refer 
to the adjoint representation, $[T^a , T^b]=f^{abc}T^c$, 
$Tr(T^a T^b)=\d^{ab}$.}.

\noindent
In the case of flat space--time, with a 
metric $\eta_{\mu\nu}$,
the action \equ{top-bf} can be rewritten as
\eq
S_{inv}= \frac{1}{4}\in \e^{\mu\nu\rho\sigma} 
F^a_{\mu\nu}B^a_{\rho\sigma} \ ,
\eqn{flat-bf}
where $\e^{\mu\nu\rho\sigma}$ denotes the totally antisymmetric tensor
of rank four.

\noindent
The action \equ{flat-bf} possesses two kinds of invariances, given by
\eqa
\delta^{(1)}A^a_\mu \= -(D_\mu \theta)^a = 
- (\6_\mu \theta^a + f^{abc}A^b_\mu \theta^c) \ , \non
\delta^{(1)}B^a_{\mu\nu} \= f^{abc}\theta^b B^c_{\mu\nu} \ , 
\eqan{sym1}
and
\eqa
\delta^{(2)}A^a_\mu \= 0 \ , \non
\delta^{(2)}B^a_{\mu\nu} \= -(D_\mu \varphi_\nu - D_\nu \varphi_\mu)^a \ , 
\eqan{sym2}
with $\theta^a$ and $\varphi^a_\mu$ as local parameters for the two 
symmetries.
Remark, that the second symmetry contains zero modes~\cite{mieg1,alvi1,dia1}, 
which we will take
into account in the next subsection for the general case of a curved 
space--time.

\subsection{The BF model in curved space--time}

>From now on, we are discussing the BF model on an arbitrary
four--manifold, endowed with an Euclidean metric $g_{\mu\nu}$.
Rewriting \equ{top-bf} in components one obtains for the invariant 
classical action in curved space--time
\eq
S_{inv}= \frac{1}{4}\in \e^{\mu\nu\rho\sigma} 
F^a_{\mu\nu}B^a_{\rho\sigma} \ ,
\eqn{curv-bf}
where the symbol $\e^{\mu\nu\rho\sigma}$ now represents, 
contrary to that one in \equ{flat-bf}, 
a totally antisymmetric tensor density with weight 1.  Furthermore,
the determinant of the metric $g=det(g_{\mu\nu})$ has weight 2 and the
volume element density $d^4x$ carries weight $-1$.  The relation
between the contravariant and covariant $\e$--tensor densities is
given by \eq \e_{\mu\nu\rho\sigma} =
g_{\mu\alpha}g_{\nu\beta}g_{\rho\gamma}g_{\sigma\delta}
\frac{1}{g}\e^{\alpha\beta\gamma\delta} \ , \eqn{epsilon} where the
weight of $\e_{\mu\nu\rho\sigma}$ is $-1$. Therefore, the action
\equ{curv-bf} is, besides the symmetries \equ{sym1} and \equ{sym2},
also invariant under diffeomorphisms with the corresponding
infinitesimal parameter $\e^\m$.

\noindent In the following we will use the BRS formalism~\cite{brs},
which requires the introduction of Faddeev--Popov ghosts $c$ and $\x$
of ghost number one for the infinitesimal parameters $\theta$ and
$\varphi$.  Collecting both symmetries in \equ{sym1} and \equ{sym2} we
get, in a first step, for the BRS transformations of the gauge field
$A^a_\mu$ and the antisymmetric tensor field $B^a_{\mu\nu}$ \eqa
sA^a_\mu \= -(D_\mu c)^a = -(\6_\mu c^a + f^{abc}A^b_\mu c^c) \ , \non
sB^a_{\mu\nu} \= -(D_\mu \xi_\nu - D_\nu \xi_\mu)^a + f^{abc} c^b
B^c_{\mu\nu} \ , \eqan{brs1} which leave the action \equ{curv-bf}
invariant.  A special care has to be taken for the $\xi$--symmetry due
to the presence of the so--called reducible symmetry in the BRS
transformations given above.  To show this fact, we rewrite the BRS
transformation of the $B$--field in terms of forms and analyze only
the part containing $\xi$, namely $s_\xi B^a=(D\xi)^a$.  This
transformation is nilpotent up to a covariant exterior derivative of
an arbitrary zero form $\phi^a$ with ghost number 2, because one has
\eq s^2_\xi B^a = (D\xi)^a = -(DD\phi)^a = -f^{abc} F^b \phi^c =
-f^{abc}\frac{\d S_{inv}}{\d B^b}\phi^c \ , \eqn{zero-mode} which
vanishes on--shell. This symmetry is said to be on--shell reducible.
Therefore, the whole set of BRS transformations is given
by\footnote{The BRS transformations of the gauge ghost $c^a$, the
vector ghost field $\xi^a_\mu$ and the scalar ghost field $\phi^a$ are
defined by the requirement of the nilpotency of $s$.}  \eqa sA^a_\mu
\= -(D_\mu c)^a = -(\6_\mu c^a + f^{abc}A^b_\mu c^c) \ , \non
sB^a_{\mu\nu} \= -(D_\mu \xi_\nu - D_\nu \xi_\mu)^a + f^{abc} c^b
B^c_{\mu\nu} \ , \non sc^a \= \half f^{abc} c^b c^c \ , \non
s\xi^a_\mu \= (D_\mu \phi)^a + f^{abc} c^b \xi^c_\mu \ , \non s\phi^a
\= f^{abc} c^b \phi^c \ .  \eqan{brs2} After some calculations one
finds \eq s^2 B^a_{\m\n} = -\half\e_{\m\n\rho\sigma}f^{abc} \frac{\d
S_{inv}}{\d B^b_{\rho\sigma}}\phi^c~~~\hbox{and}~~~ s^2 =
0~~~\hbox{for the other fields} \ .  \eqn{on-shell1}

\noindent The quantization of the model is not straightforward due to
the presence of zero modes~\cite{mieg1,alvi1,dia1} and can be
performed by using the Batalin--Vilkovisky scheme~\cite{bat1}.  In the
present work we will not follow this way, but will use another
equivalent procedure~\cite{pig3}.  Following~\cite{sor2}, the
gauge--fixing action in the Landau--type gauge, adapted to the case of
curved space--time, is given by \eqa S_{gf} \= s \in \sqrt{g} \Big[
\bar c^a g^{\mu\nu} \nabla_\mu A^a_\nu
+g^{\mu\alpha}g^{\nu\beta}\bar\xi^a_\beta\nabla_\alpha B^a_{\mu\nu}
+\bar\phi^a g^{\mu\nu} \nabla_\mu \xi^a_\nu \non
&&~~~~~~~~~~~~~~~+\,\bar\xi^a_\mu g^{\mu\nu}\nabla_\nu e^a +\bar\phi^a
\lambda^a \Big] \ , \eqan{gauge-fix} with the covariant space--time
derivative $\nabla_\mu$ defined by\footnote{Remark, that the metric
$g_{\m\n}$ is covariantly constant, \ie $\nabla_\rho g_{\m\n} = 0$.}
$\nabla_\mu X_\nu = \6_\mu X_\nu - \Gamma^\lambda_{\mu\nu} X_\lambda$,
where $\Gamma^\lambda_{\mu\nu}$ denotes the Christoffel symbol, \eq
\Gamma^\lambda_{\mu\nu} = \half g^{\lambda\rho}(\6_\m g_{\n\rho}
+\6_\n g_{\m\rho}-\6_\rho g_{\m\n}) \ , \eqn{christoffel} which is
symmetric in the lower indices assuming a torsion--free manifold.
Notice that the gauge--fixing action in \equ{gauge-fix} contains
inhomogeneous gauge conditions~\cite{leib} for the fields $B^a_{\m\n}$
and $\x^a_\m$.

\noindent The corresponding antighosts and Lagrange multiplier fields
are introduced in BRS--doublets \eqa &&s\bar c^a = b^a~~~,~~~sb^a = 0
\ , \non &&s\bar \x^a_\m = h^a_\m~~~,~~~sh^a_\m = 0 \ , \non &&s\bar
\f^a = \omega^a~~~,~~~s\omega^a = 0 \ , \non &&s e^a =
\lambda^a~~~,~~~s\lambda^a = 0 \ .  \eqan{brs-doub}

The gauge--fixing action \equ{gauge-fix} depends on the metric
explicitly and hence it has no more a topological character, but it
is still invariant under diffeomorphisms.  Furthermore, the metric
plays the role of a gauge parameter which we also let transform as a
BRS--doublet \eq sg_{\m\n}=\hat g_{\m\n}~~~,~~~s\hat g_{\m\n} = 0 \ ,
\eqn{brs-metric} in order to guarantee its non--physical
meaning~\cite{pig2}.  This is understood as the concept of extended
BRS symmetry~\cite{pig1}. Note that the BRS transformation of the
inverse of the metric is given by $sg^{\mu\nu}=-g^{\m\a}g^{\n\b}\hat
g_{\a\b}=-\hat g^{\m\n}$.

\noindent The canonical dimensions of the fields, the assigned ghost
numbers and the corresponding weights are given in Table 1.
\begin{table}[h] \renewcommand{\arraystretch}{1.2} \begin{center}
\begin{tabular}{|c|c|c|c|c|c|c|c|c|c|c|c|c|c|c|c|c|}\hline
      &$A^a_\m$ &$B^a_{\m\n}$ &$c^a$ &$\bar c^a$ &$b^a$
      &$\x^a_\m$ &$\bar\x^a_\m$ &$h^a_\m$ &$\f^a$ &$\bar\f^a$
      &$\omega^a$ &$e^a$ &$\lambda^a$ &$g_{\m\n}$ &$\hat g_{\m\n}$
      &$\sqrt g$
      \\ \hline dim &1 &2 &0 &2 &2 &1 &1 &1 &0 &2 &2 &2 &2 &0 &0 &0 \\
\hline $\phi\pi$ &0 &0 &1 &-1 &0 &1 &-1 &0 &2 &-2 &-1 &0 &1 &0 &1 &0
\\ \hline weight &0 &0 &0 &0 &0 &0 &0 &0 &0 &0 &0 &0 &0 &0 &0 &1 \\
\hline \end{tabular} \\ \vspace{0.5cm} \small{Table 1: Dimensions,
ghost numbers and weights.}  \end{center} \end{table}

\noindent Due to the gauge--fixing action term \equ{gauge-fix} the BRS
transformation of the $B$--field is no more nilpotent on--shell, since
some of the terms in the equation of motion, steaming from the
gauge--fixing part, are missing.  In order to reestablish the
nilpotency for the $B$--field one has to modify its BRS transformation
according to \eq sB^a_{\m\n}= -(D_\mu \xi_\nu - D_\nu \xi_\mu)^a +
f^{abc} c^b B^c_{\mu\nu} + \e_{\m\n\rho\sigma}f^{abc}
\sqrt{g}g^{\rho\a}g^{\sigma\b}(\6_\a\bar\x^b_\b)\f^c \ .  \eqn{brs-B}
This requires to add a further term in the gauge fixed action
$(S_{inv} + S_{gf})$ to make it invariant under the modified BRS
transformation\footnote{In \equ{gauge-fixed-action} we have performed
an integration by parts and have used the fact that the Christoffel
symbol is symmetric in the lower indices.}  \eqa S_{inv} + S_{gf} \=
\frac{1}{4}\in \e^{\mu\nu\rho\sigma} F^a_{\mu\nu}B^a_{\rho\sigma} - s
\in \sqrt{g} \Big[ g^{\mu\nu} (\6_\mu \bar c^a) A^a_\nu
+g^{\mu\alpha}g^{\nu\beta}(\6_\a\bar\xi^a_\b) B^a_{\mu\nu} \non
&+&g^{\mu\nu} (\6_\mu\bar\phi^a) \xi^a_\nu -g^{\mu\nu}\bar\xi^a_\mu
\6_\nu e^a -\bar\phi^a \lambda^a \Big] \non &-&\half\in
f^{abc}\e^{\m\n\rho\sigma}
(\6_\m\bar\x^a_\n)(\6_\rho\bar\x^b_\sigma)\f^c \ ,
\eqan{gauge-fixed-action} with \eq s(S_{inv}+S_{gf}) = 0 \ .
\eqn{invariance-action} Remark, that the last term in
\equ{gauge-fixed-action} does not disturb the topological character of
the theory.

\noindent The BRS transformations of the fields introduced so far
read: \eqa sA^a_\mu \= -(D_\mu c)^a = -(\6_\mu c^a + f^{abc}A^b_\mu
c^c) \ , \non sB^a_{\mu\nu} \= -(D_\mu \xi_\nu - D_\nu \xi_\mu)^a +
f^{abc} c^b B^c_{\mu\nu} + \e_{\m\n\rho\sigma}f^{abc}
\sqrt{g}g^{\rho\a}g^{\sigma\b}(\6_\a\bar\x^b_\b)\f^c \ , \non sc^a \=
\half f^{abc} c^b c^c \ , \non s\bar c^a \= b^a \ , \hspace{4.1cm}
sb^a = 0 \ , \non s\xi^a_\mu \= (D_\mu \phi)^a + f^{abc} c^b \xi^c_\mu
\ , \non s\bar \x^a_\m \= h^a_\m \ , \hspace{4cm} sh^a_\m = 0 \ , \non
s\bar \f^a \= \omega^a \ , \hspace{4cm} s\omega^a = 0 \ , \non s\phi^a
\= f^{abc} c^b \phi^c \ , \non s e^a \= \lambda^a \ , \hspace{4cm}
s\lambda^a = 0 \ , \non s g_{\m\n} \= \hat g_{\m\n} \ , \hspace{3.8cm}
s\hat g_{\m\n} = 0 \ .  \eqan{brs-set} These transformations are
nilpotent on--shell: \eq s^2 B^a_{\m\n} =
-\half\e_{\m\n\rho\sigma}f^{abc} \frac{\d (S_{inv}+S_{gf})}{\d
B^b_{\rho\sigma}}\phi^c~~~\hbox{and}~~~ s^2 = 0~~~\hbox{for the other
fields} \ .  \eqn{inv1}

\subsection{Supersymmetry--like transformations}

Besides the BRS symmetry and the invariance under diffeomorphisms, the
action could possess a further supersymmetric--like invariance
given by\footnote{In order to get fermionic generators we assign to
the the parameter of the diffeomorphisms $\e^\m$ ghost number 1 and to
the parameter of the supersymmetry--like transformations $\tau^\m$
ghost number 2.}  \eqa \d^S_{(\tau)}A^a_\mu \=
-\e_{\m\n\rho\sigma}\tau^\n \sqrt g
g^{\rho\a}g^{\sigma\b}\6_\a\bar\xi^a_\b \ , \non
\d^S_{(\tau)}B^a_{\m\n} \= -\e_{\m\n\rho\sigma}\tau^\rho \sqrt g
g^{\sigma\a}\6_\a\bar c^a \ , \non \d^S_{(\tau)}c^a \= -\tau^\mu
A^a_\mu \ , \non \d^S_{(\tau)}\bar c^a \= 0 \ , \non \d^S_{(\tau)}b^a
\= \LL_\tau\bar c^a = \tau^\mu\6_\mu\bar c^a \ , \non
\d^S_{(\tau)}\x^a_\m \= \tau^\n B^a_{\m\n} \ , \non
\d^S_{(\tau)}\bar\x^a_\m \= -g_{\m\n}\tau^\n \bar\f^a \ , \non
\d^S_{(\tau)}h^a_\m \= \LL_\tau\bar\x^a_\m + s(g_{\m\n}\tau^\n
\bar\f^a) = \tau^\nu\6_\nu \bar\x^a_\mu + (\6_\mu\tau^\nu)
\bar\x^a_\nu + s(g_{\m\n}\tau^\n \bar\f^a) \ , \non \d^S_{(\tau)}\f^a
\= \tau^\m \x^a_\m \ , \non \d^S_{(\tau)}\bar \f^a \= 0 \ , \non
\d^S_{(\tau)}\omega^a \= \LL_\tau\bar \f^a = \tau^\mu\6_\mu\bar \f^a \
, \non \d^S_{(\tau)}e^a \= 0 \ , \non \d^S_{(\tau)}\lambda^a \=
\LL_\tau e^a = \tau^\mu\6_\mu e^a \ , \non \d^S_{(\tau)}g_{\m\n} \= 0
\ , \non \d^S_{(\tau)}\hat g_{\m\n} \= \LL_\tau g_{\m\n} =
\tau^\rho\6_\rho g_{\mu\nu} + (\6_\mu\tau^\rho) g_{\rho\nu} +
(\6_\nu\tau^\rho) g_{\mu\rho} \ , \eqan{susy} with the corresponding
infinitesimal parameter $\tau^\m$ and the Lie derivative $\LL_\tau$.
The resultant algebra between the BRS operator $s$, the generator of
diffeomorphisms $\d^D_{(\e)}$ and the generator of
superdiffeomorphisms $\d^S_{(\tau)}$ closes on--shell \eqa &&\lac s ,
s \rac = 0~+~\hbox{equations of motion} \ , \nonumber \\[2mm] &&\lac s
, \d^S_{(\tau)} \rac = \LL_\tau~+~ \hbox{equations of motion} \ ,
\nonumber \\[2mm] &&\lac \d^S_{(\tau)} , \d^S_{(\tau)} \rac = 0 \ ,
\eqan{algebra} whereby the gauge fixed action $(S_{inv}+S_{gf})$ obeys
the following symmetries: \eqa s(S_{inv}+S_{gf}) =
\d^D_{(\e)}(S_{inv}+S_{gf}) = \d^S_{(\tau)}(S_{inv}+S_{gf}) = 0 \ .
\eqan{sym-action}

\noindent At this stage we have to make some comments about the
algebra concerning the parameter $\tau^\mu$ of the susy--like
transformations.  Contrary to the case of flat space--time, where one
has instead of $\tau^\m$ a constant parameter for the translations,
the algebra in the present case does not close {\em a priori}. In
particular, for the antisymmetric tensor field $B^a_{\m\n}$ one has
\eq \lac s , \d^S_{(\tau)} \rac B^a_{\m\n} = \LL_\tau B^a_{\m\n}
+\e_{\m\n\rho\sigma} \tau^\rho \frac{\d(S_{inv}+S_{gf})}{\d
A^a_\sigma} -\e_{\m\n\rho\sigma} f^{abc} \sqrt g g^{\rho\a} (\nabla_\a
\tau^\sigma) \bar\f^b \f^c \ .  \eqn{algebra-B} In order to implement
a closed (at least on--shell) algebra, the last term in
\equ{algebra-B}, which is quadratic in the quantum fields, has to
vanish, since it cannot be absorbed in the equation of motion. Hence,
we require \eq g^{\rho\a} (\nabla_\a \tau^\sigma) - g^{\sigma\a}
(\nabla_\a \tau^\rho) = 0 \ .  \eqn{sol} Therefore, the sysy--like
symmetry is only realizable on manifolds where \equ{sol} has a
solution. This guarantees that the algebra closes on--shell on the Lie
derivative.  In particular, a solution of \equ{sol} is given by \eq
\tau^\m = g^{\m\n} \6_\n \Lambda \ .  \eqn{condition} For completeness
we remark that also the closure of $\big\{\d^S,\d^S\big\}$ is
disturbed by a term of this kind \eq \lac \d^S_{(\tau)} ,
\d^S_{(\tau')} \rac A^a_\m = \e_{\m\n\rho\sigma} \sqrt g g^{\rho\a}
\bar\f^a (\tau'^\n \nabla_\a \tau^\sigma +\tau^\n \nabla_\a
\tau'^\sigma) \ .  \eqn{nil-a} Furthermore, when the susy--like
operator $\d^S_{(\tau)}$ acts on the gauge fixed action we get the
breaking \eq \d^S_{(\tau)}(S_{inv}+S_{gf}) = -s \in \Big( \sqrt g
g^{\m\a} (\nabla_\a \tau^\n) \bar\f^a B^a_{\m\n} + \e^{\m\n\rho\sigma}
g_{\m\l} (\nabla_\n \tau^\l) (\6_\rho \bar \xi^a_\sigma)\bar c^a \Big)
\ , \eqn{susy-break} which contains analogous terms. So one can see
that with the help of \equ{condition} one gets \equ{algebra} and the
last equation in \equ{sym-action}. \\ Finally, as we will explain in
the next section, the constraint \equ{sol} requires that the parameter
of diffeomorphisms $\e^\m$ must be a killing vector such that $\LL_\e
g_{\m\n}=0$. As a consequence, instead of diffeomorphism invariance we
have translation invariance with vector parameter a killing vector.

\subsection{The off--shell algebra} 

In order to translate the BRS invariance of the gauge fixed action
into a Slavnov identity, one has to couple the nonlinear parts of the
BRS transformations \equ{brs-set} to external sources, which lead to
the following metric--independent action term \eqa S_{ext} \= \in
\Big[ \half\gamma^{a\m\n}(sB^a_{\m\n})
+\Omega^{a\m}(sA^a_\m)+L^a(sc^a)+D^a(s\f^a)+\rho^{a\m}(s\x^a_\m)\Big]
\non &+&\frac{1}{8}\in f^{abc}\e_{\m\n\rho\sigma}\gamma^{a\m\n}
\gamma^{b\rho\sigma}\f^c \ , \eqan{ext-action} whereby all external
sources carry weight one and do not tansform under the BRS operator.
Let us remark, that the last additional term in the external action
\equ{ext-action} has an analogous origin as the one in the
gauge--fixing action \equ{gauge-fixed-action}. It ensures the Slavnov
identity \equ{slavnov} in presence of the (on--shell nilpotent) BRS
transformations \equ{brs-set} (see also~\cite{bat1}).

\noindent The dimensions, ghost numbers and weights of the sources are
given in Table 2.  \begin{table}[h] \renewcommand{\arraystretch}{1.2}
\begin{center} \begin{tabular}{|c|c|c|c|c|c|}\hline
      &$\gamma^{a\m\n}$ &$\Omega^{a\m}$ &$L^a$ &$D^a$ &$\rho^{a\m}$
      \\ \hline dim &2 &3 &4 &4 &3 \\ \hline $\phi\pi$ &-1 &-1 &-2 &-3
&-2 \\ \hline weight &1 &1 &1 &1 &1 \\ \hline \end{tabular} \\
\vspace{0.5cm} \small{Table 2: Dimensions, ghost numbers and weights
of the external sources.}  \end{center} \end{table}

\noindent
The complete classical action 
\eq
\Sigma=S_{inv}+S_{gf}+S_{ext}  
\eqn{action-classical}
obeys the Slavnov identity
\eq
\SS(\Sigma) = 0 \ ,
\eqn{slavnov}
where 
\eqa
\SS(\Sigma)\= \in \Big(\half\ds{\gamma^{a\m\n}}\ds{B^a_{\m\n}}
+\ds{\Omega^{a\m}}\ds{A^a_{\m}}+\ds{L^a}\ds{c^a}  
+\ds{D^a}\ds{\f^a}+\ds{\rho^{a\m}}\ds{\x^a_{\m}} \non
&+&b^a \ds{\bar c^a}+h^a_\m \ds{\bar\x^a_\m}
+\omega^a \ds{\bar \f^a}+\lambda^a \ds{e^a}
+\half \hat g_{\m\n} \ds{g_{\m\n}}\Big) \ .
\eqan{slavnov-identity}
It is straightforward to verify that the corresponding linearized 
Slavnov operator
\eqa
\SS_\Sigma \= \in \Big(\half\ds{\gamma^{a\m\n}}\dd{B^a_{\m\n}}
+\half\ds{B^a_{\m\n}}\dd{\gamma^{a\m\n}}
+\ds{\Omega^{a\m}}\dd{A^a_{\m}}+\ds{A^a_{\m}}\dd{\Omega^{a\m}} \non
&+&\ds{L^a}\dd{c^a}+\ds{c^a}\dd{L^a}  
+\ds{D^a}\dd{\f^a}+\ds{\f^a}\dd{D^a}
+\ds{\rho^{a\m}}\dd{\x^a_{\m}}+\ds{\x^a_{\m}}\dd{\rho^{a\m}} \non
&+&b^a \dd{\bar c^a}+h^a_\m \dd{\bar\x^a_\m}
+\omega^a \dd{\bar \f^a}+\lambda^a \dd{e^a}
+\half \hat g_{\m\n} \dd{g_{\m\n}}\Big)
\eqan{linear-slavnov-operator}
is nilpotent, \ie
\eq
\lac \SS_\Sigma , \SS_\Sigma \rac = 0 \ .
\eqn{nilpotence}

\noindent
At the functional level, the invariance of the classical action
under translations\footnote{See the last paragraph of this subsection.} 
can be expressed by an unbroken Ward identity
\eq
\PP_{(\e)} \Sigma = 0 \ ,
\eqn{ward-diff}
where $\PP_{(\e)}$ denotes the corresponding Ward operator
\eq
\PP_{(\e)} = \in \sum_{\varphi} \big( 
\LL_\e \varphi \big) \dd{\varphi} \ ,
\eqn{ward-operator-diff}
for all fields $\varphi$. Of course $\LL_\e g_{\m\n}=0$, which is the 
Killing condition (see below, (\ref{alkjfh})).

\noindent
Concerning the invariance under the rigid susy--like transformations,
the related Ward operator $\VV^S_{(\tau)}$ is given 
by\footnote{Due to the presence of the external sources the operator
$\d^S_{(\tau)}$ is modified.}
\eqa
\VV^S_{(\tau)} \= \in \Big[ 
-\e_{\m\n\rho\sigma}\tau^\n (\sqrt g   
g^{\rho\a}g^{\sigma\b}\6_\a\bar\xi^a_\b
+\half\gamma^{a\rho\sigma}) \dd{A^a_\mu} \non
&-&\half\e_{\m\n\rho\sigma}\tau^\rho (\sqrt g
g^{\sigma\a}\6_\a\bar c^a+\Omega^{a\sigma}) \dd{B^a_{\m\n}}  
-\tau^\mu A^a_\mu \dd{c^a} 
+(\LL_\tau\bar c^a) \dd{b^a}  
+\tau^\n B^a_{\m\n} \dd{\x^a_\m} \non 
&-&g_{\m\n}\tau^\n \bar\f^a \dd{\bar\x^a_\m}  
+(\LL_\tau\bar\x^a_\m 
+ s(g_{\m\n}\tau^\n \bar\f^a)) \dd{h^a_\m} 
+\tau^\m \x^a_\m \dd{\f^a} 
+(\LL_\tau\bar \f^a)\dd{\omega^a} 
+(\LL_\tau e^a)\dd{\lambda^a} \non
&-& \tau^\mu D^a \dd{\rho^{a\m}}
-\tau^\mu L^a \dd{\Omega^{a\m}}
-\tau^\mu \rho^{a\n} \dd{\gamma^{a\m\n}}
\Big] \ .
\eqan{ward-operator-susy}
After tedious calculations the corresponding Ward identity takes the form
\eq
\VV^S_{(\tau)} \Sigma = \Delta^{cl}_{(\tau)} \ ,
\eqn{ward-susy}
where the breaking writes as
\eqa
\Delta^{cl}_{(\tau)} \= \in \Big[ -\half\gamma^{a\m\n}\LL_\tau B^a_{\m\n}
-\Omega^{a\m}\LL_\tau A^a_{\m}+L^a\LL_\tau c^a-D^a\LL_\tau \f^a
+\rho^{a\m}\LL_\tau \x^a_{\m} \non
&-&\e_{\m\n\rho\sigma}\Omega^{a\m} \tau^\n 
s(\sqrt g g^{\rho\a}g^{\sigma\b}\6_\a\bar\x^a_\b)
-\half\e_{\m\n\rho\sigma}\gamma^{a\m\n} \tau^\rho 
s(\sqrt g g^{\sigma\a}\6_\a\bar c^a) \Big] \ .
\eqan{classical-breaking}
The breaking is linear in the quantum fields and therefore 
harmless in the context of the renormalization procedure~\cite{pig3}.

\noindent
It is straightforward to verify that the classical action 
\equ{action-classical} fulfills, besides the gauge--fixing conditions,
\eqa
\ds{b^a} \= \6_\m (\sqrt g g^{\m\n} A^a_\n) \ , \non
\ds{h^a_\m} \= -\6_\n (\sqrt g g^{\m\a}g^{\n\b} B^a_{\a\b}) 
+\sqrt g g^{\m\n} \6_\n e^a \ , \non
\ds{\omega^a} \= \6_\m (\sqrt g g^{\m\n} \x^a_\n)
+\sqrt g \lambda^a \ , \non
\ds{\lambda^a} \= -\6_\m (\sqrt g g^{\m\n} \bar\x^a_\n) 
-s(\sqrt g \bar\f^a)  \ ,
\eqan{gauge-conditions}
also the corresponding antighost equations,
\eqa
\ds{\bar c^a} 
+\6_\m\Big( \sqrt g g^{\m\n} \ds{\Omega^{a\n}} \Big) 
\= -\6_\m\Big( s(\sqrt g g^{\m\n}) A^a_\n \Big) \ , \non
\ds{\bar \x^a_\m} 
-\6_\n\Big( \sqrt g g^{\m\a}g^{\n\b} \ds{\gamma^{a\a\b}} \Big) 
\= \6_\n\Big( s(\sqrt g g^{\m\a}g^{\n\b}) B^a_{\a\b} \Big)
-s(\sqrt g g^{\m\n} \6_\n e^a) \ , \non
\ds{\bar \f^a} 
-\6_\m\Big( \sqrt g g^{\m\n} \ds{\rho^{a\n}} \Big) 
\=\6_\m\Big( s(\sqrt g g^{\m\n}) \x^a_\n \Big)+s(\sqrt g \lambda^a) \ , \non
\ds{e^a} \=
-\6_\m\Big( s(\sqrt g g^{\m\n}) \bar \x^a_\n  
+\sqrt g g^{\m\n} h^a_\n\Big) \ , 
\eqan{antighost}
which one usually obtains by (anti-)commuting the gauge conditions 
\equ{gauge-conditions} with the Slavnov identity \equ{slavnov}.

\noindent
Finally, the action \equ{action-classical} obeys a further integrated 
constraint, namely the ghost equation
\eq
\GG^a\Sigma = \Delta^a \ ,
\eqn{ghost-equation}
where the integrated ghost operator is given by
\eq
\GG^a = \in \Big( \dd{\f^a}-f^{abc}\bar\f^b\dd{b^c} \Big) \ ,
\eqn{ghost-operator}
and
\eq
\Delta^a = \in f^{abc} \Big( \half\e_{\m\n\rho\sigma} \gamma^{b\m\n}
\sqrt g g^{\rho\a} g^{\sigma\b} \6_\a \bar\x^c_\b + D^b c^c
+\rho^{b\m} A^c_\m + \frac{1}{8}\e_{\m\n\rho\sigma} \gamma^{b\m\n}
\gamma^{c\rho\sigma} \Big) \ .
\eqn{ghost-breaking}

\noindent
As a conclusion of this section we display the complete nonlinear algebra
generated by all the operators defined above, with $\Gamma$ an 
arbitrary functional depending on the fields of the model,
\eqa
\SS_\Gamma \SS(\Gamma) \= 0 \ , \non
\SS_\Gamma \PP_{(\e)} \Gamma + \PP_{(\e)}\SS(\Gamma) \= 0 \ , \non
\lac \PP_{(\e)} , \PP_{(\e')} \rac \Gamma \= 
-\PP_{(\lac\e,\e'\rac)} \Gamma = 0 \ , \non
\SS_\Gamma (\VV^S_{(\tau)}\Gamma-\Delta^{cl}_{(\tau)}) 
+\VV^S_{(\tau)}\SS(\Gamma) \= \PP_{(\tau)} \Gamma \ , \non
\lac \VV^S_{(\tau)} , \VV^S_{(\tau')} \rac \Gamma \= 0 \ , \non
\PP_{(\e)} (\VV^S_{(\tau)}\Gamma-\Delta^{cl}_{(\tau)}) 
+\VV^S_{(\tau)} \PP_{(\e)}\Gamma \=
-\VV^S_{(\lc\e,\tau\rc)}\Gamma = 0 \ , 
\eqan{nonlinear-algebra1}
\eqa
\GG^a \SS(\Gamma) - \SS_\Gamma (\GG^a \Gamma-\Delta^a) \= 
\FF^a \Gamma-\Theta^a \ , \non
\GG^a \PP_{(\e)}\Gamma-\PP_{(\e)}(\GG^a\Gamma-\Delta^a)\=0 \ , \non
\GG^a (\VV^S_{(\tau)}\Gamma-\Delta^{cl}_{(\tau)})
-\VV^S_{(\tau)}(\GG^a\Gamma-\Delta^a)\=0 \ , \non
\GG^a (\GG^b\Gamma-\Delta^b)-\GG^b (\GG^a\Gamma-\Delta^a) \= 0 \ , \non
\FF^a \SS(\Gamma) + \SS_\Gamma (\FF^a \Gamma-\Theta^a) \= 0 \ , \non
\FF^a \PP_{(\e)}\Gamma+\PP_{(\e)}(\FF^a\Gamma-\Theta^a)\=0 \ , \non
\FF^a (\VV^S_{(\tau)}\Gamma-\Delta^{cl}_{(\tau)})
+\VV^S_{(\tau)}(\FF^a\Gamma-\Theta^a)\=0 \ , \non
\FF^a (\GG^b\Gamma-\Delta^b)-\GG^b (\FF^a\Gamma-\Theta^a) \= 0 \ , \non
\FF^a (\FF^b\Gamma-\Theta^b)+\FF^b (\FF^a\Gamma-\Theta^a) \= 0 \ , 
\eqan{nonlinear-algebra2}
where
\eqa
\FF^a \= \in f^{abc} \Big( -\half\e_{\m\n\rho\sigma}
(\sqrt g g^{\m\a} g^{\n\b} \6_\a \bar\x^b_\b
+\half\gamma^{b\m\n})\dd{B^c_{\rho\sigma}}
+\rho^{b\m}\dd{\O^{c\m}} \non
&-&D^b\dd{L^c}-c^b\dd{\f^c}
-A^b_\m\dd{\x^c_\m}-\bar\f^b\dd{\bar c^c}
+\o^b\dd{b^c}\Big) \ ,
\eqan{f-operator}
with
\eq
\Theta^a = \in f^{abc} \half\e_{\m\n\rho\sigma} \gamma^{b\m\n}
s(\sqrt g g^{\rho\a} g^{\sigma\b} \6_\a \bar\x^c_\b)  \ .
\eqn{f-breaking}

\noindent
If the functional $\S$ is a solution of the Slavnov identity
and of the Ward identities of translations and rigid supersymmetry
the off--shell algebra \equ{nonlinear-algebra1} reduces to the linear 
algebra
\eqa
\SS_\Sigma\SS_\Sigma  \= 0 \ , \non
\lac \SS_\Sigma , \PP_{(\e)} \rac \= 0 \ , \non
\lac \PP_{(\e)} , \PP_{(\e')} \rac \= -\PP_{(\lac\e,\e'\rac)} = 0 
\ , \non
\lac \SS_\Sigma , \VV^S_{(\tau)} \rac \= \PP_{(\tau)} \ , \non
\lac \PP_{(\e)} , \VV^S_{(\tau)} \rac \= -\VV^S_{(\lc\e,\tau\rc)} =0 \ 
, \non
\lac \VV^S_{(\tau)} , \VV^S_{(\tau')} \rac \= 0 \ ,
\eqan{off-shell-algebra}
with the graded Lie brackets 
\eqa
 \{\e,\e'\}^\m \= \LL_{\e}\e'^\m 
= \e^\n\6_\n\e'^\m + \e'^\n\6_\n\e^\m \ , \non[2mm]
 [\e,\tau ]^\m \= \LL_{\e}\tau^\m 
= \e^\n\6_\n\tau^\m - \tau^\n\6_\n\e^\m \ .
\eqan{lie-def}
Let us give some remarks concerning the above results.
First, to get the Ward operator for the rigid susy--like transformations
on the right hand side of the fifth identity in \equ{off-shell-algebra}
the vector parameter $\lbrack \tau, \ \e \rbrack^\m$ must obey the 
constraint \equ{sol}. This requirement leads exactly to a further 
constraint, indeed:
\eq
\LL_\e g_{\m\n}= 0,
\eqn{alkjfh}
which means that $\e^\m$ is a Killing vector.
On the other hand, from the fourth identity in \equ{off-shell-algebra}
and the requirement above \equ{alkjfh} we conclude that also the
vector $\tau^\m$ has also to fulfill the Killing condition 
\equ{alkjfh}. To summarize, we have 
\eq
\ba{rcl}
\LL_\tau g_{\m\n} &=& \nabla_\m \tau_\n + \nabla_\n \tau_\m = 0, \\
\pa_\m \tau_\n - \pa_\n \tau_\m &=& 
\nabla_\m \tau_\n - \nabla_\n \tau_\m = 0,
\ea
\eqn{myxnbcv}
both equations imply that the vector $\tau^\m$ has to be covariantly
constant,
\eq
\nabla_\m \tau_\n = 0.
\eqn{ksfjh}
As a consequence the graded Lie brackets of two covariantly constant
vectors is zero, therefore the results in \equ{off-shell-algebra}. \\
As expected from the case of flat space--time~\cite{sor2}, the
rigid supersymmetry anticommuted with the BRS transformations yield
translations. In our case, however, the curved manifold possesses
a covariantly constant vector.

\noindent
Finally, let us remark that the classical action obeys a further
invariance, namely the rigid gauge invariance
\eq
\HH^a_{rig.} \Sigma = 0 \ ,
\eqn{rigid-1}
with the corresponding Ward operator
\eq
\HH^a_{rig.} = \sum_{\varphi} \in f^{abc} \varphi^b \dd{\varphi^c} \ ,
\eqn{rigid-2}
where $\varphi$ stands for all fields.

\section{Stability}

Till now we have been concentrated on the classical analysis of the model 
and its symmetries.
In this section, we will discuss the problem
of stability of the theory, which can be formulated as a
cohomology problem. 
By stability we mean that the most general counterterm provides
a redefinition of the fields and/or
a renormalization of the parameters of the theory which are already
present at the classical level.
Let us explicitly consider the perturbed action 
\begin{equation}
\label{per}
\Sigma^{\prime} = \Sigma + \Delta  \ ,
\end{equation}
where $\Sigma$ is the original action (\ref{action-classical}) and
$\S^\prime$ is an arbitrary functional which satisfies the Slavnov 
identity (\ref{slavnov}), the Ward identities for 
translations (\ref{ward-diff}) and 
rigid susy--like transformations (\ref{ward-susy}),
as well as the gauge conditions (\ref{gauge-conditions}), 
the antighost equations
(\ref{antighost}) and the ghost equation (\ref{ghost-equation}).
The perturbation $\Delta$ is an integrated local
polynomial of dimension four and ghost number zero. 

\noindent
The consistency with the above constraints requires the quantity $\Delta$
to fulfill the following set of equations:
\begin{equation}
\label{g1}
\frac{\d \D}{\d b^a} = 0 \ ,
\end{equation}
\label{g2}
\begin{equation}
\frac{\d \D}{\d h^a_\m} = 0 \ ,
\end{equation}
\begin{equation}
\label{g3}
\frac{\d \D}{\d \o^a} = 0 \ ,
\end{equation}
\begin{equation}
\label{g4}
\frac{\d \D}{\d \l^a} = 0 \ ,
\end{equation}
\begin{equation}
\label{a1}
\frac{\d \D}{\d \bar{c}^a} + \pan \lp \sqrt{g} g^{\m\n}                        
\frac{\d \D}{\d \O^{a\m}} \rp = 0 \ ,
\end{equation}
\begin{equation}
\label{a2}
\frac{\d \D}{\d \bar{\x}^a_\mu} - \pan \lp \sqrt{g} g^{\m\a}g^{\n\b}
\frac{\d \D}{\d \g^{a\a\b}} \rp = 0 \ ,
\end{equation}
\begin{equation}
\label{a3}
\frac{\d \D}{\d \bar{\f}^a} - \pam \lp \sqrt{g} g^{\m\n} 
\frac{\d \D}{\d \r^{a\n}} \rp = 0  \ ,
\end{equation}
\begin{equation}
\label{a4}
\frac{\d \D}{\d e^a} = 0 \ ,
\end{equation}
\begin{equation}
\label{w1}
\SS_\S \D = 0 \ ,
\end{equation}
\begin{equation}
\label{w2}
\VV^S_{(\tau)} \D = 0 \ ,
\end{equation}
\begin{equation}
\label{w3}
\PP_{(\e)} \D = 0 \ ,
\end{equation}
\eq
\label{gh1}
\int d^4x \frac{\d \D}{\d \f^a} = 0 \ .
\end{equation}
The equations (\ref{g1})--(\ref{g4}) and the equation (\ref{a4})
imply that the perturbation $\D$ does not depend on the fields 
$b^a$, $h^a_\m$, $\o^a$, $\l^a$ and $e^a$, whereas the equations
(\ref{a1})--(\ref{a3}) imply that the fields $(\bar{c}^a, \O^{a\m})$,
$(\bar{\x}^a_\m, \g^{a\m\n})$, $(\bar{\f}^a, \r^{a\m})$ can appear in
$\D$ only through the following combinations:
\eqa
\tilde{\O}^{a\m}\= \O^{a\m} + \sqrt{g}g^{\m\n}\pan \bar{c}^a \ , \non
\tilde{\g}^{a\m\n}\= \g^{a\m\n} + \sqrt{g} g^{\m\a}g^{\n\b}
(\pa_\a \bar{\x}^a_\b - \pa_\b \bar{\x}^a_\a) \ , \non
\tilde{\r}^{a\m}\= \r^{a\m} - \sqrt{g}g^{\m\n}\pan \bar{\f}^a  \ .
\eqan{redef}
The redefinitions of the external sources \equ{redef} 
imply that $\D$ is independent of the antighosts
\eqa
\frac{\d \D}{\d \bar{c}^a} \= 0 \ , ~~~
\frac{\d \D}{\d \bar{\x}^a_\mu} = 0 \ , ~~~
\frac{\d \D}{\d \bar{\f}^a} = 0  \ .
\eqan{antigh}
This means that our problem reduces to finding all possible $\D$'s such that
\eq
\D = \D(A^a_\m,c^a,B^a_{\m\n},\x^a_\m,\f^a,\tilde \g^{a\m\n},\tilde \O^{a\m}
,\tilde \r^{a\m},L^a,D^a,g_{\m\n},\hat g_{\m\n}) \ .
\end{equation}
Equations (\ref{w1})--(\ref{w3}) can be collected \cite{bbbcd} 
to produce a single cohomology problem
\begin{equation}
\label{co1}
\d \D = 0 \ .
\end{equation}
The operator $\d$ is given by 
\begin{equation}
\label{d1}
\d = \SS_\S + \PP_{(\e)} + \VV^S_{(\tau)} + \int d^4x (-\tau^\m) 
      \frac{\d}{\d \e^\m} \ .
\end{equation}
At this point let us remark that due to the redefinitions of the external 
sources \equ{redef} the Slavnov operator and the Ward operator of 
susy--like transformations are now given by\footnote{Furthermore, the Ward 
operators are already restricted to the actual field content.}
\eqa
\SS_\Sigma \!\!\=\!\!\in\Big(\half\ds{\tilde\gamma^{a\m\n}}\dd{B^a_{\m\n}}
+\half\ds{B^a_{\m\n}}\dd{\tilde\gamma^{a\m\n}}
+\ds{\tilde\Omega^{a\m}}\dd{A^a_{\m}}
+\ds{A^a_{\m}}\dd{\tilde\Omega^{a\m}} 
+\ds{L^a}\dd{c^a} \non
&+&\ds{c^a}\dd{L^a}  
+\ds{D^a}\dd{\f^a}+\ds{\f^a}\dd{D^a}
+\ds{\tilde\rho^{a\m}}\dd{\x^a_{\m}}
+\ds{\x^a_{\m}}\dd{\tilde\rho^{a\m}} 
+\half \hat g_{\m\n} \dd{g_{\m\n}}\Big) \ ,
\eqan{linear-slavnov-operator-mo}
\eqa
\VV^S_{(\tau)} \= \in \Big( 
-\half\e_{\m\n\rho\sigma}\tau^\n\tilde\gamma^{a\rho\sigma}\dd{A^a_\mu} 
-\half\e_{\m\n\rho\sigma}\tau^\rho\tilde\Omega^{a\sigma}\dd{B^a_{\m\n}}  
-\tau^\mu A^a_\mu \dd{c^a} 
+\tau^\n B^a_{\m\n} \dd{\x^a_\m} \non 
&+&\tau^\m \x^a_\m \dd{\f^a} 
 -\tau^\mu D^a \dd{\tilde\rho^{a\m}}
-\tau^\mu L^a \dd{\tilde\Omega^{a\m}} - 
\half(\tau^\mu \tilde\rho^{a\n}-\tau^\nu \tilde\rho^{a\m}) 
\dd{\tilde\gamma^{a\m\n}} \Big) \ .
\eqan{ward-operator-susy-mo}

\noindent
The main property of the above constructed operator $\d$
is its nilpotency
\begin{equation}
\label{ni1}
\d^2 = 0 \ .
\end{equation}
Thus, one can easily check that the cohomology problem (\ref{co1}) possesses 
solutions of the form $\delta = \delta \hat \Delta $. These are 
called trivial solutions because the nilpotency of $\delta$
implies that any expression of the form $\delta \hat \Delta$ 
is automatically a solution of (\ref{co1}).

\noindent
In the following we will call {\it cohomology of $\delta$} the space of all
solutions of (\ref{co1}) modulo trivial solutions.
In other words, we are looking for $\D = \D_c +\d \hat \D$, where
$\D_c$ is $\d$--closed ($\d \D_c=0$) but not trivial 
($\D_c \not= \d \tilde \D$).
We therefore introduce an operator ${\cal N}$, the filtering operator
\begin{equation}
\label{fil}
{\cal N} = \int d^4 x \sum_{f} f \frac{\delta}{\delta f} \ ,
\end{equation} 
where $f$ stands for all fields on which $\Delta$ depends.
Here, we have assigned to each field homogeneity degree 1.
The operator ${\cal N}$ induces a decomposition of $\delta$
according to
\begin{equation}
\label{dec}
\delta = \delta_0 + \delta_1 \ .
\end{equation}
The operator $\delta_0$ in (\ref{dec}) has the property that it does
not increase the homogeneity degree when it acts on a field polynomial,
whereas $\delta_1$ increases the homogeneity degree by 1.
Due to the nilpotency of $\delta$ one has also
\begin{equation}
\label{xxx}
\delta_0^2 = \lbrace \delta_0 , \delta_1 \rbrace = \d_1^2 =0 \ .
\end{equation}
An obvious identity which follows from (\ref{dec}) and 
(\ref{co1}) reads 
\begin{equation}
\label{000}
\delta_0 \Delta = 0 \ .
\end{equation}
Due to the nilpotency of $\delta_0$ (\ref{xxx}), the above equation 
(\ref{000}) defines a further cohomology problem. 

\noindent
At this stage we will use the important result \cite{dixon,brandt} 
given by the theorem, which states that
{\it the cohomology of the operator $\delta$ is isomorphic to a 
subspace of the cohomology of the operator $\delta_0$.} 

\noindent
Next, we will solve the cohomology of $\d_0$, which is easier to solve
than the cohomology of $\d$ and by using the theorem mentioned above
we will determine the solution of $\d\D=0$.
The action of the operator $\d_0$ on the fields is explicitly given by
\eqa
\d_0 A^a_\m \= - \pam c^a \ , \non
\d_0 c^a \= 0 \ , \non
\d_0 B^a_{\m\n} \= -\pam \x^a_\n + \pan \x^a_\m \ , \non  
\d_0 \x^a_\m \= \pam \f^a \ , \non
\d_0 \f^a \= 0 \ , \non
\d_0 g_{\m\n}\=\hat g_{\m\n} \ , \non 
\d_0\hat g_{\m\n}\=0 \ , \non
\d_0 \tilde \O^{a\mu} \= \frac{1}{2} \e^{\m\n\r\s} \pan B^a_{\r\s} \ , \non
\d_0 \tilde \g^{a\m\n} \= \e^{\m\n\r\s}\pa_\r A^a_\s \ , \non
\d_0 L^a \= - \pam \tilde \O^{a\m} \ , \non 
\d_0 D^a \= -\pam\tilde \r^{a\m} \ , \non
\d_0 \tilde\r^{a\m} \= \pan \tilde \g^{a\m\n} \ , \non
\d_0 \e^\mu \= - \tau^\mu \ , \non
\d_0 \tau^\mu \= 0 \ .
\eqan{delta-null}
The first remark we make is that $g_{\m\n}$ and $\hat g_{\m\n}$ 
and also $\e^\mu$ and $\tau^\mu$
transform as $\d_0$--doublets, therefore both pairs of fields are
out of the cohomology \cite{brandt}, implying that $\D_c$ is independent 
of $g_{\m\n}$, $\hat g_{\m\n}$, $\e^\mu$ and $\tau^\mu$.  
In order to have a more compact notation 
we switch to the language of forms where $d$ represents the 
nilpotent ($d^2=0$) exterior derivative, given explicitly by 
$d = \dxm \pam$. The gauge field $A^a_\m$ and the ghost field $\x^a_\m$
are represented by the one forms $A^a = A^a_\m \dxm$ and 
$\x^a = \x^a_\m \dxm$, and the antisymmetric tensor field $B^a_{\m\n}$ 
by the two form $B^a = \frac{1}{2}B^a_{\m\n}\dxm\dxn$. 
The ghosts $c^a$ and $\f^a$
are scalar fields or equivalently zero forms.

\noindent
>From the quantities $\tilde \g^{a\m\n}, \tilde \r^{a\m}, 
\tilde \O^{a\m},L^a$ 
and $D^a$ we construct the following dual forms:
\eqa
\as\tilde\g^a \=\frac{1}{2!}\e_{\m\n\r\s}\tilde\g^{a\m\n} \dxr dx^\s \ , \non
\as\tilde\r^a \=\frac{1}{3!}
\e_{\m\n\r\s}\tilde\r^{a\m}\dxn\dxr dx^\s \ , \non
\as\,\tilde\O^a \=\frac{1}{3!}
\e_{\m\n\r\s}\tilde\O^{a\m}\dxn\dxr dx^\s \ , \non
\as L^a \=\frac{1}{4!}\e_{\m\n\r\s}L^a \dxm\dxn\dxr dx^\s \ , \non
\as D^a \=\frac{1}{4!}\e_{\m\n\r\s}D^a \dxm\dxn\dxr dx^\s \ .
\eqan{sjkdfhg}
In terms of forms the nilpotent operator $\d_0$ reads\footnote{The 
functional derivatives with respect to differential forms has to be 
understood as follows:
\eq
\frac{\delta S}{\delta f} = X ~~~\hbox{if}~~~ S=\int_{\MM} f\,X \ , \nonumber
\eqn{foot}
where $f$ and $X$ are general differential forms.}
\eqa
\d_0 \= \int_\MM \Big( dc^a \frac{\d}{\d A^a} +
d\x^a \frac{\d}{\d B^a} +
d\f^a \frac{\d}{\d \x^a} +
2dA^a \frac{\d}{\d \as\tilde \g^a} +
dB^a \frac{\d}{\d \as\,\tilde \O^a} +  
d \as\,\tilde \O^a \frac{\d}{\d \as L^a} - \non
&-&d \as\tilde \r^a \frac{\d}{\d \as D^a} -     
\half d \as\tilde \g^a \frac{\d}{\d \as\tilde \r^a} \Big) + 
\int d^4x \Big( \half \hat g_{\m\n}\frac{\d}{\d g_{\m\n}} 
- \tau^\mu \dd{\e^\mu}\Big) \ .
\eqan{delta0}
For $\D_c$ being an integrated polynomial of form degree four and 
ghost number zero we can write $\D_c = \int_\MM \o^0_4$, where $\o^p_q$ is 
a field polynomial of form degree $q$ and ghost number $p$. 
Due to Stocks´ theorem and to (\ref{000}),
we note that $\d_0 \D_c = \int_\MM \d_0 \o^0_4=0$, 
which implies the following result
\eq
\d_0 \o^0_4 + d \o^1_3 = 0 \ .
\end{equation}
Using the algebraic Poincar\'e lemma~\cite{brandt} 
and the fact that $\d_0$ and
$d$ anticommute, $\lbrace \d_0,\! d \rbrace =0$, we derive
the following tower of descent equations
\eqa
\d_0 \o^0_4 + d \o^1_3 \= 0 \ , \non
\d_0 \o^1_3 + d \o^2_2 \= 0 \ , \non
\d_0 \o^2_2 + d \o^3_1 \= 0 \ , \non
\d_0 \o^3_1 + d \o^4_0 \= 0 \ , \non
\d_0 \o^4_0  \= 0 \ .
\eqan{tower}
The only possible expression for $\o^4_0$, restricted by its form degree 
and ghost number, is given by
\eq
\o^4_0 = u \f^a \f^a + v f^{abc} c^a c^b \f^c \ ,
\end{equation}
where $u$ and $v$ are constant coefficients. In order to solve the tower
of descent equations \equ{tower} we decompose the exterior 
derivative~\cite{sps} according to
\eq
\big[\bar\delta,\delta_0 \big] = d \ , ~~~ \big[\bar\delta,d \big] = 0 \ ,
\eqn{BRS_COMM}
with the operator $\bar\delta$ given by
\eqa
\bar\delta\=-\as\tilde\gamma^a\pp{A^a}-A^a\pp{c^a}
-3\as\,\tilde\Omega^a\pp{B^a}-2B^a\pp{\xi^a}
-\xi^a\pp{\f^a} \non
&+&6\as\tilde\rho^a\pp{\as\tilde\gamma^a}
-4\as L^a\pp{\as\,\tilde\Omega^a}
+4\as D^a\pp{\as\tilde\rho^a}  \ .
\eqan{DECOMP}
With the help of the operator \equ{DECOMP} one finds
\eqa
\o^0_4 \= \bar\delta\bar\delta\bar\delta\bar\delta \, \o^4_0  \non
\= u \Big(\as L^a \f^a + \as\,\tilde\O^a\x^a + \frac{1}{2} B^aB^a \Big) \non
&+& v f^{abc} \Big( \as D^a c^b \f^c - \as\tilde \r^a A^b \f^c -
\as\tilde \r^a c^b \x^c 
+ \frac{1}{8} \as\tilde \g^a \as\tilde \g^b \f^c + \non
&+& \half\as\tilde \g^a A^b \x^c + \frac{1}{2} A^a A^b B^c + 
\half\as\tilde \g^ac^bB^c + A^ac^b\as\,\tilde \O^c 
+ \frac{1}{2}c^ac^b\as L^c \Big) \ .
\eqan{omega-4}
After some calculations one can show the invariance of
$\int_\MM \o^0_4$ under the whole operator $\d$ of (\ref{d1}) 
\eq
\d \int_\MM \o^0_4 = 0 \ .
\end{equation}
The general solution of the cohomology problem \equ{co1}
has the form
\eq
\label{ct}
\D = \D_c + \d \hat \D = \int_\MM \CC^0_4 + \d \hat \D \ . 
\end{equation}
The nontrivial solution of the $\d$--cohomology, $\int_\MM \CC^0_4$,
must not necessarily be equal to the nontrivial solution 
of the $\d_0$--cohomology $\int_\MM \o^0_4$.
In order to analyze this situation 
completely, we construct the most general trivial solution,
restricted by the dimension, the ghost number and the weight, 
according to
\newpage
\eqa
\hat \D \= \int d^4x \Big( \a_1 \tilde \O^{a\m} A^a_\m + \a_2
\tilde \g^{a\m\n} B^a_{\m\n} + \a_3 L^a c^a + \a_4
\tilde \r^{a\m} \x^a_\m + \a_5 D^a \f^a \non 
&+&\a_6 f^{abc} \tilde \g^{a\m\n} A^b_\m A^c_\n  
+ \a_7 f^{abc} \tilde \r^{a\m} A^b_\m c^c + \a_8 \tilde \g^{a\m\n} 
\pam A^a_\n + \a_9
\tilde \r^{a\m} A^a_\n g^{\r\n} \hat g_{\r\m} \non[2mm]
&+& \a_{10} f^{abc} D^a c^b c^c 
+ \a_{11} f^{abc} \e_{\m\n\r\s} 
\tilde \g^{a\m\n} \tilde \g^{b\r\s} c^c
+ \a_{12}  D^a c^a \hat g_{\m\n} g^{\m\n} + \a_{13} \tilde \r^{a\m} 
\pam c^a  \non
&+& \a_{14} \frac{1}{\sqrt{g}} \tilde \g^{a\m\n} 
\tilde \g^{a\r\s} \hat g_{\m\r} g_{\n\s} 
+ \a_{15} \sqrt{g}\e_{\m\n\r\s} 
\tilde \g^{a\m\n} g^{\r\a}g^{\s\b} \pa_\a A^a_\b  \non
&+& \a_{16} \frac{1}{\sqrt{g}} f^{abc} g_{\m\r}g_{\n\s} 
\tilde \g^{a\m\n} \tilde \g^{b\r\s} c^c
+ \a_{17} \tilde \r^{a\m} A^a_\m g^{\r\s} \hat g_{\r\s} \non
&+& \a_{18} \sqrt{g}\e_{\m\n\r\s} f^{abc} 
\tilde \g^{a\m\n} g^{\r\a} g^{\s\b}
A^b_\a A^c_\b + \a_{19}\sqrt{g}
\e_{\m\n\r\s} g^{\r\a} g^{\s\b} \tilde \g^{a\m\n} B^a_{\a\b} \Big) \ ,
\eqan{delta-hat}
with $\a_i$, $i=1,...,19$ as constant coefficients.

\noindent
The only acceptable counterterms for the 
action (\ref{action-classical})
must be independent of the vector parameters $\e^\m$ and $\tau^\m$.
A careful analysis of the $\e^\m$-- and $\tau^\m$--dependent 
part\footnote{The technical details can be found in the appendix.} 
of $\d\hat\D$ leads to the vanishing of all 
$\a_i$ except $\a_1$, $\a_2$, $\a_3$, $\a_4$, and $\a_5$ 
which fulfill 
\eq
-\a_5 = \a_4 = -\a_3 = -2\a_2 = \a_1 \equiv \a \ .
\end{equation}
Therefore, the counterterms (\ref{ct}) reduce to
\eq
\label{ct1}
\D = \int_\MM \CC^0_4 + \a\,\SS_\Sigma \int d^4x \Big(
\tilde \O^{a\m} A^a_\m 
- \half\tilde \g^{a\m\n} B^a_{\m\n} - L^a c^a +
\tilde \r^{a\m} \x^a_\m - D^a \f^a \Big) \ .
\end{equation}

\noindent
An important fact is that the $\a$--proportional term in \equ{ct1},
after performing the $\SS_\Sigma$--operation, gives identically the
$v$--proportional part of \equ{omega-4}. This means that \equ{omega-4}
contains a trivial solution of the $\SS_\S$--cohomology, which
can be reabsorbed in the trivial counterterms.
Therefore, the complete expression for the counterterms \equ{ct} is given by
\eqa
\D \= u\int_\MM \Big(\as L^a \f^a +\as\,\tilde\O^a\x^a 
+\frac{1}{2}B^aB^a\Big) \non
&+& \a\,\SS_\Sigma \int d^4x \Big(
\tilde \O^{a\m} A^a_\m 
- \half\tilde \g^{a\m\n} B^a_{\m\n} - L^a c^a +
\tilde \r^{a\m} \x^a_\m - D^a \f^a \Big)  \ .
\eqan{ct-fin}
Now, by using the ghost equation, or more precisely the 
constraint (\ref{gh1}) we deduce the following result: 
the two constant coefficients $u$ and $\a$ present in 
the counterterms (\ref{ct-fin}) are both equal to zero. 
This means that there
is no possible deformation of the action (\ref{action-classical}), 
which is the most general local functional, 
solution of the Ward identities, the gauge 
conditions, the antighost equations as well as the ghost equation.

\noindent
On the other hand, the result of this section implies that at the 
quantum level (if there are no anomalies)
the four dimensional antisymmetric tensor field model does not admit any
renormalizations (renormalization of the coupling constant or of 
the fields). 
In this case the theory is said to be finite. To prove
the finiteness to all orders of perturbation theory one has to 
overcome another problem: the absence of anomalies. 
This is the subject of the last section.

\section{Anomaly analysis}

In the context of renormalization theory one has to investigate 
whether the symmetries, collected in
$\delta$, are not disturbed by quantum corrections. 
If there is an anomaly, then it corresponds to $\d \S= \AA$,
where $\AA$ is an integrated local field polynomial of form degree 
four and ghost number one ($\AA= \int_\MM \o^1_4$), that fulfills 
\eq
\label{an}
\d \AA = 0 \ .
\end{equation}

\noindent
Using the same strategy as in the previous section, we derive
the following tower of descent equations
\eqa
\d_0 \o^1_4 + d \o^2_3 \= 0 \ , \non
\d_0 \o^2_3 + d \o^3_2 \= 0 \ , \non
\d_0 \o^3_2 + d \o^4_1 \= 0 \ , \non
\d_0 \o^4_1 + d \o^5_0 \= 0 \ , \non
\d_0 \o^5_0  \= 0 \ . 
\eqan{tower-anomaly}
The only possible expression for $\o^5_0$ is given by
\eq
\label{anom}
\o^5_0 = x Tr(c^5) + y Tr(c^3 \f) + z Tr(c \f^2) \ ,
\end{equation}
with $x$, $y$ and $z$ constant coefficients. 
The last two terms in \equ{anom} are $\d_0$ invariant, but not 
$\d$ invariant expressions. Since a possible anomaly has 
to be invariant under the $\d$ operation, one has to set the
coefficients $y$ and $z$ equal to zero.

\noindent
Using the decomposition operator \equ{DECOMP}, the solution of 
the descent equations is given by
\eqa
\o^1_4 \!\=\bar\d\bar\d\bar\d\bar\d\o^5_0 = \non
\=\!\!x Tr \big( 
             \as D c^4 - \as\tilde \r A c^3 -\as\tilde \r c A c^2 -
             \as\tilde \r c^2 A c - \as\tilde \r c^3 A + 
             \frac{1}{4}\as\tilde \g^2 c^3 + 
             \frac{1}{4}\as\tilde \g c \as\tilde \g c^2 +
             \frac{1}{2}\as\tilde \g A^2 c^2 + \non
      &+&    \frac{1}{2}\as\tilde \g AcAc + 
             \frac{1}{2}\as\tilde \g Ac^2A +  
             \frac{1}{2}\as\tilde \g cA^2c + 
             \frac{1}{2}\as\tilde\g cAcA + 
             \frac{1}{2}\as\tilde\g c^2A^2 + A^4c 
             \big) \ ,
\eqan{deform}
which belongs not only to the cohomology of $\d_0$ modulo $d$ but also
to the cohomolgy of $\d$ modulo $d$, \ie $\d\o^1_4+d\o^2_3=0$. 
This implies that the anomaly candidate,
\ie solution of (\ref{an}), is nothing else but
\eqa
\AA \= x Tr \int_\MM \big(  
             \as D c^4 - \as\tilde \r A c^3 -\as\tilde \r c A c^2 -
             \as\tilde \r c^2 A c - \as\tilde \r c^3 A + 
             \frac{1}{4}\as\tilde \g^2 c^3 + 
             \frac{1}{4}\as\tilde \g c \as\tilde \g c^2 + \non
      &+&    \frac{1}{2}\as\tilde \g A^2 c^2 + 
             \frac{1}{2}\as\tilde \g AcAc + 
             \frac{1}{2}\as\tilde \g Ac^2A +  
             \frac{1}{2}\as\tilde \g cA^2c + 
             \frac{1}{2}\as\tilde\g cAcA +  \non
      &+&    \frac{1}{2}\as\tilde\g c^2A^2 + A^4c 
             \big) \ .
\eqan{anc}
As argued in \cite{sor2}, the anomaly canditate $\AA$ disappears due to 
the fact that all the fields considered so far take values in 
the adjoint representation of the gauge group. In this case
the totally symmetric tensor defined by the symmetrized trace
of the generators of the gauge group, $d^{abc}=\half Tr(T^a\{T^b,T^c\})$,
which is present in the trace of \equ{anc}, vanishes.

\noindent
Therefore, the most general solution of $\d \AA=0$ is a $\d$--exact quantity
given by $\AA= \d \hat \AA$. This particularly means that
the Slavnov identity, the translations and the rigid susy--like 
transformations Ward identities are anomaly free, thus
can be promoted to the quantum level.
Using standard arguments \cite{pig3} one can easily show that 
the constraints
(\ref{g1})--(\ref{a4}) are anomaly free, hence valid to all orders
of perturbation theory. Concerning the ghost equation (\ref{xxx}), 
it can also
be proven to hold at the quantum level. The proof may be carried out by
following the lines of \cite{bps}. 

\noindent
In this section we have showed that the four dimensional antisymmetric
tensor field model in a curved background, admitting a covariantly constant
vector, is anomaly free. Therefore, due to the results of the
previous section, it is finite to all orders of perturbation theory.

\section{Conclusion}

We have shown in great details that the four dimensional antisymmetric
tensor field model in a curved space--time is finite to all orders of
perturbation theory. The proof was performed by an extensive use of the 
algebraic renormalization procedure, which does not depend on a particular
regularization scheme such as the dimensional regularization or the BPHZ 
regularization. But, unfortunately, the use of the algebraic 
renormalization scheme requires the existence of a possible regularization 
{\em a priori}. 
This fact limits our quantum analysis to be only valid in the case 
of a curved, topologically trivial and asymptotically flat manifolds
admitting covariantly constant vector.

On the other hand, we have seen that the role played by the symmetry
under the susy--like transformations was decisive in reducing the 
counterterms (\ref{ct}) to take the simpler form (\ref{ct1}), 
which was forbidden by the ghost equation. 
This symmetry only exists, as it was
shown in Section \ref{sec1}, on manifolds where the equation \equ{sol}
has a solution. Remember also, that we have considered manifolds where the 
torsion vanishes.
These are all the restrictions on the manifolds where our quantum 
analysis holds.

\section*{Appendix: Analysis of the trivial counterterms}

\setcounter{equation}{0}
\renewcommand{\theequation}{A.\arabic{equation}}

We devote this appendix to give all the superdiffeomorphism parameter 
dependent field polynomials appearing in the trivial counterterms 
constructed in (\ref{ct}). 
With 
\eq
\VV^S_{(\tau)}\hat\Delta=\sum_{i=1}^{19}X_i 
\eqn{xor}
these polynomials read as 
\eqa
X_1 &=& \int d^4x ~\a_1~ \Big[ - L^a \tau^\m A^a_\m +
\half\e_{\m\n\r\s} \tilde \O^{a\m}\tau^\n \tilde \g^{a\r\s} 
\Big] \ , \\
X_2 &=& \int d^4x ~\a_2~ \Big[ 
2 \tilde \r^{a\m}\tau^\n B^a_{\m\n} + 
\e_{\m\n\r\s} \tilde \g^{a\m\n}\tau^\r \tilde \O^{a\s}
\Big] \ , \\
X_3 &=& \int d^4x ~\a_3~ \Big[ 
- L^a \tau^\m A^a_\m
\Big] \ , \\
X_4 &=& \int d^4x ~\a_4~ \Big[ 
- D^a \tau^\m \x^a_\m + \tilde \r^{a\m} \tau^\n B^a_{\m\n}
\Big] \ , \\
X_5 &=& \int d^4x ~\a_5~ \Big[ 
- D^a \tau^\m \x^a_\m
\Big] \ , \\
X_6 &=& \int d^4x ~\a_6~ f^{abc} \Big[ 
2 \tau^\n \tilde \r^{a\m} 
A^b_\m A^c_\n + 
\e_{\m\b\r\s} \tau^\b \tilde \g^{a\m\n} \tilde \g^{b\r\s} A^c_\n
\Big] \ , \\
X_7 &=& \int d^4x ~\a_7~ f^{abc} \Big[
- \tau^\m D^a A^b_\m c^c - 
\half\e_{\m\n\r\s} \tau^\n \tilde \r^{a\m} \tilde \g^{b\r\s} c^c -
\tau^\n \tilde \r^{a\m} A^b_\m A^c_\n
\Big] \ , \\
X_8 &=& \int d^4x ~\a_8~ \Big[
- (\tau^\m \tilde \r^{a\n} - \tau^\n \tilde \r^{a\m})\pam A^a_\n 
- \half\e_{\n\r\s\b} \tau^\r 
(\partial_\mu \tilde \g^{a\m\n}) \tilde \g^{a\s\b}
\Big] \ , \\
X_9 &=& \int d^4x ~\a_9~ \Big[
- \tau^\m D^a A^a_\n g^{\r\n} \hat g_{\r\m} - \half\e_{\n\a\b\d} \tau^\a
\tilde \r^{a\m} \tilde \g^{a\b\d} g^{\r\n} \hat g_{\m\r} \Big] \ , \\
X_{10} &=& \int d^4x ~\a_{10}~\Big[
2 f^{abc} D^a \tau^\m A^b_\m c^c
\Big] \ , \\
X_{11} &=& \int d^4x ~\a_{11}~ f^{abc} \Big[
- 4 \e_{\m\n\r\s} \tau^\m \tilde \r^{a\n} \tilde \g^{b\r\s} c^c 
- \e_{\m\n\r\s} \tilde \g^{b\m\n} \tilde \g^{b\r\s} \tau^\a A^c_\a 
\Big] \ , \\
X_{12} &=& \int d^4x ~\a_{12}~ \Big[
D^a \tau^\r A^a_\r \hat g_{\m\n} g^{\m\n} \Big] \ , \\
X_{13} &=& \int d^4x ~\a_{13}~ \Big[ 
- \tau^\m D^a \pam c^a + (\partial_\mu \tilde \r^{a\m}) \tau^\n A^a_\n
\Big] \ , \\
X_{14} &=& \int d^4x \frac{\a_{14}}{\sqrt{g}} \Big[ 
2 \tau^\n \tilde \r^{a\m} 
\tilde \g^{a\r\s} \hat g_{\m\r} g_{\n\s} 
+ 2 \tilde \g^{a\m\n}
 \tau^\r \tilde \r^{a\s} 
\hat g_{\m\r} g_{\n\s} \Big] \ , \\
X_{15} &=& \int d^4x ~\a_{15}~\sqrt{g} \Big[ 
2\e_{\m\n\r\s} 
\tau^\n \tilde \r^{a\m}  (\pa_\a A^a_\b)
g^{\r\a} g^{\s\b} - \non 
&-& \half
\e_{\m\n\r\s} (\partial_\a \tilde \g^{a\m\n}) g^{\r\a} g^{\s\b} 
\e_{\b\d\l\k} \tau^\d \tilde \g^{a\l\k}
\Big] \ , \\
X_{16} &=& \int d^4x \frac{\a_{16}}{\sqrt{g}} f^{abc} \Big[
4\tau^\n \tilde \r^{a\m} \tilde \g^{b\r\s} c^c g_{\m\r} g_{\n\s}
- \tilde \g^{a\m\n} \tilde \g^{b\r\s} \tau^\a A^c_\a g_{\m\r}
g_{\n\s}
\Big] \ , \\
X_{17} &=& \int d^4x ~\a_{17}~ \Big[
-D^a \tau^\m A^a_\m g^{\r\s} \hat g_{\r\s} - \half\e_{\m\n\r\s} \tau^\n
\tilde \r^{a\m} \tilde \g^{a\r\s} g^{\a\b} \hat g_{\a\b} \Big] \ , \\
X_{18} &=& \int d^4x ~\a_{18}~\sqrt{g} f^{abc} \Big[
2\e_{\m\n\r\s}\tau^\n \tilde\r^{a\m} A^b_\a A^c_\b g^{\r\a}g^{\s\b} + \non
&+&
\e_{\m\n\r\s} \tilde \g^{a\m\n} \e_{\a\l\d\k} \tau^\l \tilde \g^{b\d\k} 
A^c_\b g^{\r\a} g^{\s\b}
\Big] \ , \\
X_{19} &=& \int d^4x ~\a_{19}~\sqrt{g} \Big[
2\e_{\m\n\r\s} \tau^\n \tilde \r^{a\m} B^a_{\a\b} g^{\r\a} g^{\s\b} + 
\e_{\m\n\r\s} \e_{\a\b\l\k} \tau^\l \tilde \g^{a\m\n} \tilde \O^{a\k}
g^{\r\a} g^{\s\b}
\Big] \ .  
\eqan{xxkjgf}
The sum of the above constructed polynomials ($\sum_i X_i$) has to vanish, 
otherwise we will get the participation of the vector 
parameter $\tau^\m$ in the expression of the counterterms (\ref{ct}), 
a fact which is not desirable.
By direct computation one can convince himself that the only possible 
solution of the constraint
\eq
\sum^{19}_{i=1} X_i = 0
\eqn{yyy}
is that all the $\a_i$ vanish for $6 \leq i \leq 19$. The remaining $\a_i$
have to obey the following equalities
\eq
-\a_5 = \a_4 = -\a_3 = -2\a_2 = \a_1 \equiv \a \ .
\end{equation} 
In this way we could reduce the trivial counterterms given in (\ref{ct}) 
to the more simpler expression (\ref{ct-fin}).



\end{document}